\documentclass{elsart}
\usepackage{graphicx}
\usepackage{amssymb}
\begin{document}

\begin{frontmatter}
\title{Determination of the $s$-wave pion-nucleon threshold scattering parameters from the results of experiments on pionic hydrogen}
\author[GC]{G.C. Oades},
\author[GR]{G. Rasche},
\author[WS]{W.S. Woolcock},
\author[EM]{E. Matsinos{$^*$}},
\author[AG]{A. Gashi}
\address[GC]{Institute of Physics and Astronomy, Aarhus University, DK-8000 Aarhus C, Denmark}
\address[GR]{Institut f\"{u}r Theoretische Physik der Universit\"{a}t, Winterthurerstrasse 190, CH-8057 Z\"{u}rich, Switzerland}
\address[WS]{Department of Theoretical Physics, IAS, The Australian National University, Canberra, ACT 0200, Australia}
\address[EM]{Varian Medical Systems Imaging Laboratory GmbH, T\"{a}fernstrasse 7, CH-5405 Baden-D\"{a}ttwil, Switzerland}
\address[AG]{Mediscope AG, Alfred Escher-Str. 27, CH-8002 Z\"{u}rich, Switzerland}

\begin{abstract}
We give the conversion equations which lead from experimental values of the $3p\rightarrow 1s$ transition energy in pionic hydrogen and the total width 
of the $1s$ level to values of the $s$-wave threshold scattering parameters for the processes $\pi^-p\rightarrow\pi^-p$ and $\pi^-p\rightarrow\pi^0n$ 
respectively. Using a three-channel potential model, we then calculate the electromagnetic corrections to these quantities, which remove the 
ef\mbox{}fects of the Coulomb interaction, the external mass dif\mbox{}ferences and the presence of the $\gamma n$ channel. We give the $s$-wave 
scattering parameters obtained from the present experimental data and these electromagnetic corrections. F\mbox{}inally we discuss the implications for 
isospin invariance.
\end{abstract}
{$^*$}{Corresponding author. E-mail address: evangelos.matsinos@varian.com. Tel.: +41 56 2030460. Fax: +41 56 2030405}
\end{frontmatter}

\section{Introduction}

Accurate values of the $s$-wave scattering parameters $a_{cc}$ and $a_{c0}$, for the processes $\pi^-p\rightarrow\pi^-p$ and $\pi^-p\rightarrow\pi^0n$ 
at the $\pi^-p$ threshold, can be obtained from measurements of an $np\rightarrow 1s$ transition energy and the width of the $1s$ level in pionic 
hydrogen (H) via a ref\mbox{}inement of the DGBT formulae \cite{DGBT}. As the most accurate experimental data available are for the $3p\rightarrow 1s$ 
transition energy, we will consider only this case in the present paper. The most recent fully published experimental values of these quantities are given 
in Ref.~\cite{A}. A successor experiment \cite{B} is nearing completion, and preliminary results from this experiment for the transition energy \cite{C} 
and for the width \cite{U} are available. In Section 2 of this paper we shall give a detailed derivation of the relevant conversion equations 
which connect the experimental data with the scattering parameters, paying particular attention to the sources of uncertainty and making clear which 
pion-nucleon ($\pi N$) interactions contribute to the parameters. We use these conversion equations to obtain the values of $a_{cc}$ and $a_{c0}$ from the 
results in Refs.~\cite{C,U}. These equations can also be used for the f\mbox{}inal results of the experiment \cite{B} and for those of any future 
experiment.

The conversion equations just mentioned are based on the formalism of Ref.~\cite{G}. It was the f\mbox{}irst work to generalise the systematic derivation 
of the DGBT formulae for a one-channel system \cite{Trueman,Lambert} to a multichannel system, and to apply the results to pionic H. This systematic 
derivation is based on the general principles of scattering theory and on the analytic continuation of the matrix $\mathbf{K}$ (def\mbox{}ined later in 
Eq.~(\ref{eq:c})) into the complex $s$-plane, to those (complex) values of $s$ which correspond to the levels of pionic H. All this is independent of any 
phenomenological model of the dynamics of the system. The formalism of Ref.~\cite{G} enables one to extract directly and unambiguously from the 
experimental data the values of precisely def\mbox{}ined threshold scattering parameters $a_{cc}$ and $a_{c0}$, whose errors are dominated by the 
experimental errors of these atomic data. Then, as a second quite distinct step, a dynamical calculation of the electromagnetic (EM) corrections to these 
quantities is needed if we are to obtain hadronic scattering parameters.

In Section 3 we shall describe our calculation of the electromagnetic (EM) corrections to the threshold parameters, which uses a three-channel potential 
model to remove the ef\mbox{}fects of the Coulomb interaction, the external mass dif\mbox{}ferences and the presence of the $\gamma n$ channel. We shall 
also give the values, together with estimated errors, of the corrected scattering parameters (denoted by $\tilde{a}_{cc}$ and $\tilde{a}_{c0}$) which 
result from $a_{cc}$ and $a_{c0}$. In Section 4 we shall set our calculation of the EM corrections in the context of previously published calculations, 
and attempt to summarise our present understanding of these corrections. We shall see that, while the experimental data from pionic H yield extremely 
accurate values of $a_{cc}$ and $a_{c0}$, the calculation of the EM corrections needed to obtain hadronic quantities is subject to a great deal of 
uncertainty.

In Section 5 we shall use the value of the $s$-wave scattering length $\tilde{a}_{+}$ for $\pi^{+}p$ elastic scattering, obtained from a recent 
phase-shift analysis (PSA) \cite{D} of the data up to a laboratory pion kinetic energy ($T_{\pi}$) of $100$ MeV, which employed the EM corrections of 
Ref.~\cite{E}, also calculated using a potential model. To ensure that the two calculations of EM corrections were consistent, we adopted, for the 
hadronic potentials in the calculation described in Section 3, the same forms as were used for the corrections given in Ref.~\cite{E}. We shall show that 
this value of $\tilde{a}_{+}$ shows a large and signif\mbox{}icant dif\mbox{}ference from the value of $\tilde{a}_{cc}+ \sqrt{2}\tilde{a}_{c0}$, 
calculated from the most recent values of $\tilde{a}_{cc}$ and $\tilde{a}_{c0}$ given in Section 3.

This reinforces the evidence for the violation of isospin invariance given in Ref.~\cite{D}, from the inability of the phase shifts, obtained from the 
analysis of $\pi^{\pm}p$ elastic scattering data alone, to reproduce the data on $\pi^-p$ charge exchange scattering. As discussed in Ref.~\cite{D}, at 
least part of this violation of isospin invariance has to be attributed to the presence of residual EM ef\mbox{}fects in the phase shifts obtained from 
the experimental data after the application of our corrections. This is because the EM corrections calculated using a potential model are only partial 
(stage 1) corrections; further (stage 2) corrections are needed to remove these residual EM ef\mbox{}fects.

We do not include in the present paper the information from the accurate measurement of a transition energy in pionic deuterium (D) \cite{Deuterium}. 
There is an extensive literature on this subject \cite{ELT,ELW,MRR}. The most recent of these papers \cite{MRR} concludes tentatively that, when stage 2 
corrections (calculated using chiral perturbation theory at leading order) are taken into account, the atomic data from pionic H and pionic D taken by 
themselves satisfy isospin invariance at the hadronic level. However, Ref.~\cite{MRR} emphasises that an improved measurement of a transition energy in 
pionic D is needed, as is a major ef\mbox{}fort on the theoretical side, to extract a scattering parameter with better precision. This complex subject lies 
outside the scope of this paper. Our aim here is to present what can be learned directly from the experimental data for pionic H and to relate the 
results consistently to those obtained from our PSA \cite{D}.

\section{Conversion formulae}

The f\mbox{}irst piece of experimental data from pionic H is the $3p\rightarrow 1s$ transition energy $W(3p\rightarrow 1s)$. The positions of the $3p$ 
and $1s$ levels relative to the $\pi^-p$ threshold we call $W(3p)$ and $W(1s)$ respectively. For the numerical calculations we will use always the most 
recent values of the physical constants \cite{F}; errors will be given only if they are relevant for our purpose.

Nonrelativistically the point charge Coulomb potential $V^{pc}$ contributes \\$-359.438$ eV to $W(3p)$. The shift due to vacuum polarisation is $-0.011$ 
eV \cite{I}. Furthermore the $3p$ level is split by the hyperf\mbox{}ine interaction and the relativistic ef\mbox{}fects. The former is discussed on page 
413 of Ref.~\cite{G}, while the latter is given by Eqs.~(9,10) of Ref.~\cite{H}. For $j = 3/2$ these two ef\mbox{}fects add to $-0.0013$ eV, for $j = 1/2$ 
to $-0.0055$ eV. With the present experimental resolution of $W(3p\rightarrow1s)$ this splitting cannot be seen. We therefore take a weighted average of 
$-0.003$ eV, with an uncertainty of $0.002$ eV, as the contribution of these ef\mbox{}fects to $W(3p)$. The shift in the $3p$ level due to the hadronic 
interaction, including the ef\mbox{}fect of the extended charge distributions of $\pi^-$ and $p$, is negligible. Thus
\[\hspace{2.7cm}
W(3p) = -359.452(2)\, \mathrm{eV}\, .
\]
Since in the transition $3p\rightarrow 1s$ the kinetic recoil energy of the $\pi^-$ atom in the c.m. frame is $0.004$ eV, we have
\begin{equation} \hspace{2cm}
W(1s) = - W(3p\rightarrow1s) - 359.456(2)\,\mathrm{eV}\,.
\label{eq:a}
\end{equation}

Nonrelativistically $V^{pc}$ contributes $-3234.945$ eV to $W(1s)$. The relativistic shift for an inf\mbox{}initely heavy proton is $-0.215$ eV. This 
comes either from Ref.~\cite{H} or from integration of the Klein-Gordon equation \cite{I}. In addition, both these references give a relativistic recoil 
shift of $+0.037$ eV and a contribution of $+0.010$ eV due to the magnetic moment of the proton. Ref.~\cite{I} also quotes a contribution of $-0.018$ eV 
for vacuum polarisation at $\it{O}(\alpha^3)$ and a vertex correction of $+0.007$ eV. The potential $V^{vp}$ for vacuum polarisation at $\it{O}(\alpha^2)$ 
is the Uehling potential, modif\mbox{}ied at short distances due to the f\mbox{}inite extension of the charge distributions of $\pi^-$ and $p$. Adding 
$V^{vp}$ to $V^{pc}$, our calculations give a shift of $-3.244$ eV. The f\mbox{}inite extension of the charge distributions also modif\mbox{}ies $V^{pc}$ 
itself through the addition of a potential
\[
V^{ext} = -\alpha/r (er\!f(r/c) - 1),\hspace{.4cm}\mathrm{with}\hspace{0.5cm} c^2 = \frac{2}{3}(<r^2>_\pi +  <r^2>_p) \, .
\]
The rms charge radii of $\pi^-$ and $p$ were taken from Refs.~\cite{K,L}. The addition of $V^{ext}$ to $V^{pc}$ gives a further shift of $+0.100$ eV. The 
$1s$ position arising from the purely electromagnetic contributions considered so far is $-3238.268$ eV. Nearly all the dif\mbox{}ference between this value and 
the widely used ``total electromagnetic binding energy'' in Table 1 of Ref.~\cite{I} comes from the change in the value of $-\frac{1}{2} \alpha^2 m_c$, due 
to the use of the most recent value \cite{F} of the mass $\mu_c$ of the $\pi^-$.

There are sizeable contributions to $W(1s)$ from the interference of $V^{vp}$ and $V^{ext}$ with the hadronic interaction $\widetilde{V}$ (see Section 4; 
$\widetilde{V}$ as well as the other potentials are matrices in a coupled-channel calculation). We determined this shift, as well as the width, directly 
in a subthreshold calculation with mass dif\mbox{}ferences included. To this end we evaluated the shift and the width due to $\widetilde{V}$, for the two 
cases $V^{pc}$ + $V^{vp}$ + $V^{ext}$ + $\widetilde{V}$ and $V^{pc}$ + $\widetilde{V}$. It turned out that, when the direct contributions of $V^{vp}$ and 
$V^{ext}$ ($-3.244$ eV and $+0.100$ eV respectively) are accounted for, the shift due to the interference of ($V^{vp} + V^{ext}$) with $\widetilde{V}$ is 
numerically negligible. The error, estimated by varying the range of $\widetilde{V}$, is $0.003$ eV. The cancellation of the two interference 
contributions due to $V^{vp}$ and $V^{ext}$ was already noted implicitly in Ref.~\cite{I}. The result for the width will be given later.

The shift of the purely electromagnetic $1s$ position given above, due to the interference of $\widetilde{V}$ with $V^{pc}$ only, we call $\Delta W$. 
Thus, from the result in the preceding paragraph, we have
\[\hspace{3cm}
W(1s) = - 3238.268(8) \, \mathrm{eV} + \Delta W  \, ,\]
and, using Eq.~(\ref{eq:a}), 
\begin{equation}\hspace{2.7cm}
\Delta W = - W(3p\rightarrow1s) + 2878.812(8)\, \mathrm{eV}\,.
\label{eq:b}
\end{equation}
Note that, for practical purposes, the shift $\Delta W$ def\mbox{}ined in this way is equal to the quantity $\epsilon_{1s}$ def\mbox{}ined in 
Refs.~\cite{A,C,I}. This is because, though we have taken account of the short-range ef\mbox{}fects fully, by including the interference of $V^{pc}$ with 
($V^{vp} + V^{ext}$) in the calculation of the shift and width (rather than as a correction to the scattering parameters), the ef\mbox{}fect is negligible 
in the case of the shift. In the last two equations we have given the uncertainty explicitly. It is dominated by the uncertainty in $\mu_c$.

The hadronic interaction couples the $\pi^-p$ channel to the $\pi^0n$ and $\gamma n$ channels. As well as creating the shift $\Delta W$, it also makes 
the $1s$ level unstable, with a width $\Gamma$. Compared to $\Gamma$, the hadronic and electromagnetic widths of the $3p$ level can be neglected 
numerically. The process then is a three-body decay of the well def\mbox{}ined $3p$ level, going via the unstable $1s$ level: 
$3p\rightarrow(\pi^0n)\gamma$ or $(\gamma n)\gamma$. The parentheses indicate that $(\pi^0n)$ and $(\gamma n)$ are the decay products of the intermediate 
$1s$ level.

The treatment of such a three-body decay mechanism is given in detail in Section 4 of Ref.~\cite{M}. Let $s$ be the square of the invariant mass of 
$(\pi^0n)$ (or $(\gamma n)$). It turns out that the decay amplitude is proportional to
\[\hspace{4.5cm}
(s - M_0^2 + iM_0\Gamma)^{-1}\, .
\]
Here $M_0 = \mu_c + m_p + W(1s)$, $m_p$ being the proton mass. We have neglected a possible dependence of $\Gamma$ on $s$. Using the relation between 
$s$ and the energy $E$ of the photon in the rest frame of the decaying $3p$ state, and noting that $W(3p\rightarrow1s)<<M_0$, one sees that the spectrum 
of $E$ becomes proportional to
\[\hspace{4cm}
[(E - W(3p\rightarrow1s))^2 + \frac{1}{4}\Gamma^2]^{-1}\, .
\]
Thus the measurement of this spectrum gives $W(3p\rightarrow1s)$ and $\Gamma$.

We now derive the conversion equations which give the values of the elements $a_{cc}$ and $a_{c0}$ of $\bf{a}$, the $s$-wave scattering matrix for 
the ($\pi^-p$,$\pi^0n$,$\gamma n$) system at the $\pi^-p$ threshold, in terms of $W(3p\rightarrow1s)$ and $\Gamma$. We use $c$, $0$ and $\gamma$ to 
denote the three coupled channels and $q_c$, $q_0$ and $k$ to denote the channel momenta in the c.m. frame. The matrix $\bf{a}$ is the value of 
$\bf{K}$ at the $\pi^-p$ threshold $q_c =0$. $\bf{K}$ is def\mbox{}ined in terms of the unitary matrix $\bf{S}$ by
\[ \hspace{3.5cm}
\mathbf{t} = -i(\mathbf{S - 1_3})(\mathbf{S + 1_3})^{-1}    \, ,
\]

\begin{equation} 
\mathbf{K}^{-1} = \left( \begin{array}{ccc}
C_{0}(\eta) & 0 & 0 \\
0           & 1 & 0 \\
0           & 0 & 1 \\
  \end{array} \right)     \mathbf{Q}^{1/2} \mathbf{t}^{-1}  \mathbf{Q}^{1/2} \left( \begin{array}{ccc}
C_{0}(\eta) & 0 & 0 \\
0           & 1 & 0 \\
0           & 0 & 1 \\
  \end{array} \right)
 +
 \left( \begin{array}{ccc}
-\beta h(\eta) & 0 & 0 \\
0              & 0 & 0 \\
0              & 0 & 0 \\
\end{array} \right) \, .
\label{eq:c}
\end{equation}
Equation (\ref{eq:c}) takes account of the low energy behaviour of $\bf t$ induced by the long range part of the Coulomb interaction. The quantities 
appearing in Eq.~(\ref{eq:c}) are
\begin{equation} 
\mathbf{Q} =  \left( \begin{array}{ccc}
q_c & 0 & 0 \\
0 & q_0 & 0 \\
0 & 0 & k  \end{array} \right),\,C_0 (\eta) = \frac{2\pi\eta}{\exp(2\pi\eta)-1},\,h(\eta)=-\ln{|\eta|} +\Re{\psi(1+i\beta)},
\label{eq:d}
\end{equation}
with $\hspace{2cm} \beta= 2\alpha m_c\,\,,\,\,\eta=-\alpha m_c/q_c \,\,,\,\, m_c = \mu_cm_p/(\mu_c+m_p)\,.\hspace{2cm}$

The matrix $\bf{K}$ is real-valued for $s>(\mu_c+m_p)^2$ and can be analytically continued into the $s$-plane.

The formalism of Ref.~\cite{G} will now be used to obtain the conversion formulae. Strictly speaking, the formalism relates $\Delta W$ and $\Gamma$ to 
$K_{cc}$ and $K_{c0}$, evaluated at the position of the $1s$ level. For reasons discussed in Section 3, phase shift analysis of low energy $\pi^- p$ 
elastic scattering does not give any reliable knowledge of the range parameters for $K_{cc}$ and $K_{c0}$. However, it is suf\mbox{}f\mbox{}icient to note 
that the dif\mbox{}ferences between the values of $K_{cc}$ and $K_{c0}$ at the $\pi^- p$ threshold and at the position of the $1s$ level are expected to 
be of order $1\cdot 10^{-6}$ fm, well below the uncertainties due to the experimental error. We therefore work from now on with the matrix $\bf{a}$ 
def\mbox{}ined above. Following Ref.~\cite{G} we introduce the matrix $\bf{A}$ by
\begin{equation} \hspace{3cm}
\mathbf{A}^{-1} = \mathbf{a}^{-1} -i 
\left( \begin{array}{ccc}
0 & 0 & 0 \\
0 & q_0 & 0 \\
0 & 0 & k  \end{array} \right)\,\, ,
\label{eq:e}
\end{equation} 
where $q_0$ and $k$ are from now on the c.m. channel momenta at the $\pi^-p$ threshold $q_c = 0$. Numerically we will use $q_0 = 28.0408(96)$ MeV/c and 
$k = 129.40705(31)$ MeV/c.

From Eq.~(\ref{eq:e}) we have
\begin{equation}
a_{cc} = \Re{A}_{cc} + q_0^2a_{c0}^2a_{00} + 2q_0ka_{c0}a_{c\gamma}a_{0\gamma} + k^2a_{c\gamma}^2a_{\gamma\gamma}     \,\,\,,
\label{eq:f}
\end{equation}
\begin{equation}
a_{c0}= -|A_{c0}|(1 + q_0k a_{0\gamma}^2 +  \frac{1}{2}q_0^2a_{00}^2 -  P^{-\frac{1}{2}}q_0^{\frac{1}{2}}k^{\frac{3}{2}}a_{0\gamma}a_{\gamma\gamma} - 
\frac{1}{2}P^{-1}q_0ka_{0\gamma}^2)       \,\,.\label{eq:g}
\end{equation}

Equation (\ref{eq:f}) is straightforward; Equation (\ref{eq:g}) uses the result $a_{c\gamma} = -a_{c0}q_0^{\frac{1}{2}}k^{-\frac{1}{2}}P^{-\frac{1}{2}}$, 
where $P$ is the Panofsky ratio, whose value \cite{N} is $ 1.546(9)$. Equations (\ref{eq:f},\ref{eq:g}) give only the lowest terms in an expansion. To 
assess the magnitude of these terms we need estimates for $a_{c0}$, $a_{c\gamma}$, $a_{00}$, $a_{0\gamma}$, $a_{\gamma\gamma}$. The values of these 
quantities will be discussed in Section 3. Numerically, we f\mbox{}ind that
\begin{equation} \hspace{1cm}
a_{cc} = \Re{A}_{cc} (1 + 4.9\cdot10^{-5}) \,\,,\,\,\, a_{c0} = -|A_{c0}|(1 + 1.1\cdot10^{-6})                          \,.
\label{eq:h} 
\end{equation}
The correction terms in Eq.~(\ref{eq:h}) are negligible.

The expression for $\Delta W$ is given in Eq.~(12) of Ref.~\cite{G} with $n=1$, $l=0$. Keeping only the numerically signif\mbox{}icant terms we have 
\begin{equation}\hspace{1cm}
\frac{\Delta W}{W(1s)^{pc}} =  \Re{\epsilon_{0}}( 1 + p_1\Re{\epsilon_{0}}[1-(\frac{\Im{\epsilon}_{0}}{\Re{\epsilon_{0}}})^2] + p_2 (\Re{\epsilon_{0}})^2 ) \,,
\label{eq:i} 
\end{equation}
where 
\begin{equation}
W(1s)^{pc} = -\frac{1}{2}\alpha^2m_c\,\,,\,\, \Re{\epsilon_{0}} =4a_{cc}/B\,\,,\,\,\Im{\epsilon}_{0} =4q_0(1+P^{-1})|a_{c0}|^2/B \,,
\label{eq:j} 
\end{equation}
$B = 1/\alpha m_c$ being the lowest Bohr radius of the $\pi^- p$ atom. The numerical values are $m_c =121.49719(26)\, \mathrm{MeV}$, 
$W(1s)^{pc}=-3234.945(7)\,\mathrm{eV}$ and $B=222.56399(48)\,\mathrm{fm}$. The quantity $p_1 = 1/2(1+\gamma)= 0.78860783$, $\gamma$ being Euler's 
constant. The expression for $p_2$ is given in Ref.~\cite{G}. The term involving a poorly known ef\mbox{}fective range parameter is small, and the result 
$p_2 = -0.22$ is suf\mbox{}f\mbox{}iciently accurate for our purpose.

From the experiment \cite{B}, Ref.~\cite{C} gives $W(3p \rightarrow 1s)= 2885.928(10)$ eV. Using Eq.~(\ref{eq:b}), this results in 
$\Delta W = -7.116(13)$ eV. With this number as the starting point for the iteration of Eq.~(\ref{eq:i}), we obtain
\begin{equation}\hspace{2.5cm}
a_{cc}(\mathrm{fm}) = 0.0171704(1)(-\Delta W(\mathrm{eV})) \,\,\,.
\label{eq:k} 
\end{equation}
The error in the constant in Eq.~(\ref{eq:k}) comes from the uncertainty in $m_c$ (due almost entirely to the error in $\mu_c$) and the uncertainty in 
$\Delta W$ (via the iteration). However, it is so small that it is of no consequence and will be disregarded in what follows. Using Eq.~(\ref{eq:b}), the 
f\mbox{}irst conversion formula is
\begin{equation}\hspace{.5cm}
a_{cc}(\mathrm{fm}) = (0.0171704)[W(3p\rightarrow 1s)(\mathrm{eV}) - 2878.812(8)] \,\,\,.
\label{eq:l} 
\end{equation}
Numerically we obtain
\begin{equation}
\hspace{3.5cm} a_{cc} = 0.12218(22) \,\mathrm{fm}\,\,\,\, .
\label{eq:m}
\end{equation}
We would like to stress that the present experimental results for $W(3p\rightarrow 1s)$ and for $\Gamma$ \cite{C,U} are consistent with, but more precise 
than, the earlier ones given in Ref.~\cite{A}. The result in Eq.~(\ref{eq:m}) is by far the most accurate number in low energy $\pi N$ physics, and 
further ref\mbox{}inement will not be useful, in view of the uncertainty in $\mu_c$ and even more of the uncertainty in the EM corrections to be 
discussed later.

We come now to the measured total width $\Gamma$ for the decay of the atomic $1s$ state into both the $\pi^0 n$ and $\gamma n$ channels. The numerical 
result for the Panofsky ratio $P$ quoted earlier is used f\mbox{}irst to obtain the partial width $\Gamma_0$ for decay into $\pi^0 n$:
\begin{equation}
\hspace{3.5cm} \Gamma = (1 + P^{-1}) \Gamma_0 \,\,\,\, .
\label{eq:n}
\end{equation}

Vacuum polarisation and the f\mbox{}inite extension of the charges each have a small but signif\mbox{}icant ef\mbox{}fect on $\Gamma_0$. The subthreshold 
calculation discussed earlier in connection with the shift shows that $V^{ext}$ changes $\Gamma_0$ by the factor $0.9948$, while $V^{vp}$ changes it by 
the factor $1.0049$. The total ef\mbox{}fect of ($V^{vp} + V^{ext}$) is a change by the factor $0.9996$.

The treatment of decay widths on pages $410$ and $411$ of Ref.~\cite{G} gives
\begin{equation}
\hspace{0.8cm} \Gamma_0 =4\alpha^3 m^2_c q_0 a_{c0}^2 ( 1 + 2 p_1\Re{\epsilon_{0}} + p_1^2 |\epsilon_{0}|^2 + 2p_2 (\Re{\epsilon_{0}})^2 )\,\,,
\label{eq:p}
\end{equation}
keeping only terms which are signif\mbox{}icant numerically and replacing $|A_{c0}|^2$ by $a_{c0}^2$ (see Eq.~(\ref{eq:h})). Combining 
Eqs.~(\ref{eq:n},\ref{eq:p}) and using numerical values already given leads to the second conversion formula
\begin{equation}\hspace{.5cm}
\hspace{2cm} a_{c0}(\mathrm{fm}) = -0.19042(22) (\Gamma(\mathrm{eV})) \, ^{1/2}  \,\,\,.
\label{eq:q} 
\end{equation}
Almost all the error in Eq.~(\ref{eq:q}) arises from the uncertainty in $P$. An updated result for $\Gamma$ from the experiment \cite{B} is $0.823(19)$ 
eV \cite{U}. This leads via Eq.~(\ref{eq:q}) to
\begin{equation}
\hspace{3.5cm} a_{c0} = -0.1727(20) \,\mathrm{fm}\,\,\,\, .
\label{eq:r}
\end{equation}
All the error in Eq.~(\ref{eq:r}) comes from the experimental uncertainty on $\Gamma$; the error due to the uncertainty of the conversion constant is 
negligible. The experiment \cite{B} f\mbox{}inally hopes to achieve an accuracy of 1\% (an absolute uncertainty of $0.008$ eV), which will reduce the 
error on $a_{c0}$ to about $0.0009$ fm. Even in this case the error due to the conversion constant is only $2\cdot 10^{-5}$ fm; therefore, a more 
accurate measurement of the Panofsky ratio would not improve the accuracy of the f\mbox{}inal value of $a_{c0}$.

\section{Calculation of electromagnetic corrections}

To calculate the electromagnetic corrections to $a_{cc}$ and $a_{c0}$ due to the point-charge Coulomb interaction, the external mass dif\mbox{}ferences 
and the presence of the $\gamma n$ channel, we set up a potential model for the physical three-channel situation at the $\pi^- p$ threshold. The 
relativised Schr$\ddot{\mathrm{o}}$dinger equation (RSE) for the $s$-wave has the form
\begin{equation}
\hspace{1.5cm}[ d^2/dr^2\mathbf{1}_3 + \mathbf{Q}^2 - 2 \tilde{\mathbf{m}}( \widetilde{\mathbf{V}}+\mathbf{V}^{pc}) ]\mathbf{u}(r)= \mathbf{0}
\label{eq:s}       \,\,\, \,,     
\end{equation}
where, since we are working at the $\pi^- p$ threshold $q_c = 0$,
\[ 
\hspace{2.5cm}\mathbf{Q} = \left( \begin{array}{ccc}
0 & 0 & 0 \\
0 & q_0 & 0 \\
0 & 0 & k  \end{array} \right),
\]
with $q_0$ and $k$ having the same meaning as in Eq.~(\ref{eq:e}). The diagonal matrix $\tilde{\mathbf{m}}$ of modif\mbox{}ied reduced masses is
\begin{equation}
\hspace{2.5cm} \tilde{\mathbf{m}}=  \left( \begin{array}{ccc}
\tilde{m}_c & 0 & 0 \\
0 & \tilde{m}_0 & 0 \\
0 & 0 & \tilde{m}_\gamma  \end{array} \right)
         \,\,\, \,. 
\label{eq:t}
\end{equation}
For a two-body channel with masses $m_1$ and $m_2$, the modif\mbox{}ied reduced mass is
\[
\hspace{3cm} \tilde{m} = \frac{W^2 - m_1^2 - m_2^2}{2W}\,\, .
\]
Thus, for $ W = \mu_c + m_p$,
\begin{equation}
\tilde{m}_c =\frac{\mu_cm_p}{\mu_c+m_p}=m_c \,,\, \tilde{m}_0=\frac{(\mu_c+m_p)^2-\mu_0^2-m_n^2}{2(\mu_c+m_p)} \,\,,\, \newline
 \tilde{m}_\gamma = \frac{(\mu_c+m_p)^2-m_n^2}{2(\mu_c+m_p)}.
\label{eq:u}
\end{equation}
The reasons for the inclusion of these relativistic modif\mbox{}ications of the reduced masses for the $\pi^-p$ and $\pi^0n$ channels were discussed in 
detail in Section 3 of Ref.~\cite{P}, for the two-channel potential model calculation of the EM corrections for the analysis of low energy $\pi^- p$ 
scattering data. The modif\mbox{}ication is essential for the $\gamma n$ channel, where the standard reduced mass is zero, whereas $\tilde{m}_\gamma$ in 
Eq.~(\ref{eq:u}) is in fact just $k$.

The $3\times3$ matrix $\mathbf{V}^{pc}$ in Eq.~(\ref{eq:s}) is
\[
\hspace{2.5cm}\mathbf{V}^{pc} = \left( \begin{array}{ccc}
V^{pc} & 0 & 0 \\
0 & 0 & 0 \\
0 & 0 & 0  \end{array} \right) \,\,\,,
\]
where $V^{pc}$ is the attractive point-charge Coulomb potential. We recall that the ef\mbox{}fects of vacuum polarisation and of the f\mbox{}inite 
extension of the charges of $\pi^-$ and $p$ have already been accounted for. We write the matrix $\widetilde{\mathbf{V}}$ in the form
\noindent 
\begin{equation} 
\widetilde{\mathbf{V}}\!=  \!\!\left( \begin{array}{ccc}
\widetilde{V}_{cc} & \widetilde{V}_{c0} & V_{c\gamma} \\
\widetilde{V}_{c0} & \widetilde{V}_{00} & V_{0\gamma} \\
V_{c\gamma} & V_{0\gamma} & V_{\gamma\gamma} \end{array} \right)
         \,\,\,  \,\mathrm{and \,def\mbox{}ine \,the\, submatrix} \,\, \widetilde{\mathbf{V}}_{h}\!=\!\!\left( \begin{array}{cc}
\widetilde{V}_{cc} & \widetilde{V}_{c0}  \\
\widetilde{V}_{c0} & \widetilde{V}_{00}  \\
\end{array} \right).
\label{eq:v}
\end{equation}
We shall call $\widetilde{\mathbf{V}}_{h}$ the EM modif\mbox{}ied hadronic potential because, as explained in Section 1, it contains residual EM 
ef\mbox{}fects. There is clear evidence, to be discussed in Section 5, that $\widetilde{\mathbf{V}}_{h}$ violates isospin invariance, though only by a few 
percent. However, we have only the numbers $a_{cc}$ and $a_{c0}$ available experimentally ($a_{00}$ is inaccessible) and therefore need to express 
$\widetilde{\mathbf{V}}_{h}$ in terms of two parameters. Hence we made the approximation of treating $\widetilde{\mathbf{V}}_{h}$ as isospin invariant:
\[\hspace{2cm}
\widetilde{\mathbf{V}}_{h} =
\left( \begin{array}{cc}
\frac{2}{3}\widetilde{V}_{1/2} +\frac{1}{3}\widetilde{V}_{3/2}& \,\,\frac{\sqrt{2}}{3}(\widetilde{V}_{3/2}- \widetilde{V}_{1/2}) \\
\,\,\frac{\sqrt{2}}{3}(\widetilde{V}_{3/2}- \widetilde{V}_{1/2}) & \frac{1}{3}\widetilde{V}_{1/2} +\frac{2}{3}\widetilde{V}_{3/2} \\
\end{array} \right).
\]
We made the same approximation in Ref.~\cite{P} in order to calculate EM corrections for the analysis of low energy $\pi^- p$ scattering data. Since this 
potential is used solely for the calculation of the corrections, this approximation should have very little ef\mbox{}fect on the results. For 
$\widetilde{V}_{1/2}$ and $\widetilde{V}_{3/2}$ we assumed the radial dependence of the $s$-wave hadronic potentials used in Refs.~\cite{E,P} and left 
their strengths as parameters to be adjusted. The leading term in each potential is of gaussian form, with a range parameter which is f\mbox{}ixed at 
$1 \,\mathrm{fm}$ (a realistic value for the hadronic interaction), and two small correction terms are added.

The three potentials $V_{c\gamma} ,\, V_{0\gamma} \,\mathrm{and}\,\, V_{\gamma\gamma}$ in Eq.~(\ref{eq:v}) also need to have some assumed radial 
dependence, with their strengths left as parameters to be adjusted. A study of the analytic structure of the relevant amplitudes shows that the range 
parameter $1 \,\mathrm{fm}$ is also appropriate for these three potentials. For convenience we took the radial dependence of $V_{c\gamma} $ to be the 
same as that of $\widetilde{V}_{c0}$, and the radial dependence of $V_{0\gamma}$ and $V_{\gamma\gamma}$ to be the same as that of $\widetilde{V}_{00}$. 
In the calculation there were then f\mbox{}ive potential strengths to be adjusted, with f\mbox{}ive experimentally determined numbers used to f\mbox{}ix 
them.

In addition to the values of $a_{cc}$ and $a_{c0}$ given in Eqs.~(\ref{eq:m},\ref{eq:r}), we used experimentally determined values of 
$a_{c\gamma} ,\, a_{0\gamma} \,\mathrm{and}\,\, a_{\gamma\gamma}$. From the values of $a_{c0}$ and $P$, we obtain $a_{c\gamma} = 0.0647$ fm. The value 
of $a_{0\gamma}$ can be obtained only from an analysis of experimental data near threshold for the photoproduction process $\gamma n \rightarrow \pi^0n$. 
The results given in Ref.~\cite{Q} lead to $a_{0\gamma}=-0.0028\,\mathrm{fm}$, but no error is given; we think that $0.0006\,\mathrm{fm}$ is a realistic 
uncertainty. The value of $a_{\gamma\gamma}$ comes from the analysis of data on Compton scattering of\mbox{}f neutrons. From results in Ref.~\cite{R} we 
have extracted $a_{\gamma\gamma}=0.0011\,\mathrm{fm}$, with an uncertainty of around $20\%$. The f\mbox{}ive potential strengths were then varied until 
they reproduced the f\mbox{}ive numbers just discussed, thus f\mbox{}ixing $\widetilde{\mathbf{V}}$ for the three-channel calculation.

We now give some details of the calculation of $\mathbf{a}$ from the RSE (\ref{eq:s}). For the two-channel case (omitting the $\gamma n$ channel) the 
calculation is given fully in Section 3 of Ref.~\cite{S}; it can easily be generalised to the three-channel situation. The f\mbox{}irst step is to form 
the matrix $\mathbf{P}(r)$ by putting side by side, as its three columns, three linearly independent regular solutions $\mathbf{u}^{j}(r), j=1,2,3,$ of 
Eq.~(\ref{eq:s}). Let $R$ be a distance beyond which $\widetilde{\mathbf{V}}$ is negligible. We then def\mbox{}ine the real symmetric matrix $\mathbf{d}$ 
by
\[\hspace{3cm}
\mathbf{d}= R\tilde{\mathbf{m}}^{-1/2}\mathbf{P}^{'}(R)[\mathbf{P}(R)]^{-1}\tilde{\mathbf{m}}^{1/2}\,\,,
\]
where the prime denotes dif\mbox{}ferentiation with respect to $r$ and the matrix elements of $\tilde{\mathbf{m}}$ are given in Eq.~(\ref{eq:u}). Note 
that our present notation is dif\mbox{}ferent from that of Ref.~\cite{S}. The matrix $\mathbf{K}$ in Eq.~(64) of that reference is now $\mathbf{t}$, 
def\mbox{}ined just before Eq.~(\ref{eq:c}), and $\mathbf{A}$ in Eq.~(65) is now $\mathbf{K}$ as def\mbox{}ined in Eq.~(\ref{eq:c}). Equation (66) of 
Ref.~\cite{S}, generalised to the three-channel case and taken at threshold, gives
\begin{equation} 
R^{-1}\mathbf{a} = \begin{array}{cc}
\left( \begin{array}{ccc}
\Phi_0^* -d_{cc}\Phi_0   & -d_{c0}\Phi_0 & -d_{c\gamma}\Phi_0 \\
-d_{c0}\rho_0^{-1}\sin{\rho_0} & \,\,\cos{\rho_0}-d_{00}\rho_0^{-1}\sin{\rho_0} & -d_{0\gamma}\rho_0^{-1}\sin{\rho_0} \\
-d_{c\gamma}\rho_\gamma^{-1}\sin{\rho}_\gamma &-d_{0\gamma}\rho_\gamma^{-1} \sin{\rho_\gamma} &     \cos{\rho_\gamma}-d_{\gamma\gamma}\rho_\gamma^{-1}\sin{\rho_\gamma} \end{array} \right) \\  \times
\left( \begin{array}{ccc}
X_0^* +d_{cc}X_0   & d_{c0}X_0 & d_{c\gamma}X_0 \\
d_{c0}\cos{\rho_0} & \,\,\rho_0 \sin{\rho_0}+d_{00}\cos{\rho_0} & d_{0\gamma}\cos{\rho_0} \\
d_{c\gamma}\cos{\rho}_\gamma &d_{0\gamma}\cos{\rho_\gamma} &     \rho_\gamma \sin{\rho_\gamma}+d_{\gamma\gamma}\cos{\rho_\gamma} \end{array} \right)^{\mathbf{-1}} \end{array},
\label{eq:w}
\end{equation}
where $\rho_0=q_0R$ and $\rho_\gamma=kR$. The auxiliary Coulomb functions $\Phi_0$, $\Phi^*_0$, $X_0$ and $X^*_0$ (discussed at length in Section 2 of 
Ref.~\cite{S}) have at threshold the arguments $(- \beta R;0)$, which we have suppressed. In terms of Bessel functions they are, with $x= 2\sqrt{\beta R}$,
\begin{equation}
\Phi_0=2x^{-1}J_1(x),\,\Phi_0^*=J_0(x),\,X_0=-\frac{1}{2}\pi xY_1(x),\,X_0^*=\frac{1}{4}x^2Y_0(x).
\label{eq:x}
\end{equation}

Having given the details of the modelling of the three-channel situation, we now proceed to calculate the corrections which remove from $a_{cc}$ and 
$a_{c0}$ the ef\mbox{}fects of the point charge Coulomb potential, of the mass dif\mbox{}ferences $\mu_c-\mu_0$ and $m_n-m_p$, and of the presence of the 
$\gamma n$ channel. To obtain them, we need to model the two-channel situation with these three ef\mbox{}fects removed, using the RSE at threshold
\begin{equation}
\hspace{3cm}[ d^2/dr^2 \mathbf{1}_2  - 2 {m}_c \widetilde{\mathbf{V}}_h ]\mathbf{u}(r)= \mathbf{0} \,\,\, .
\label{eq:y}       \,\,\, \,     
\end{equation}
The appearance of $m_c$ in Eq.~(\ref{eq:y}) implies the assignment of the mass $\mu_c$ to both $\pi^-$ and $\pi^0$, and the mass $m_p$ to both $p$ and 
$n$. This choice is made in most studies of the low energy $\pi N$ system and was discussed fully in Section 5.2 of Ref.~\cite{D}. Thus the scattering 
parameters $\tilde{a}_{cc}$ and $\tilde{a}_{c0}$ which result from Eq.~(\ref{eq:y}) still contain residual EM contributions.

By integrating Eq.~(\ref{eq:y}) we obtain the $2\times2$ matrix $\tilde{\mathbf{a}}$; the numerical values for the relevant matrix elements are
\begin{equation}\hspace{2.5cm}
\tilde{a}_{cc}= 0.12140\,\mathrm{fm}\,\,,\,\,\,\,  \tilde{a}_{c0}=-0.1757\, \mathrm{fm}\,\,\,.
\label{eq:z}
\end{equation}
We now def\mbox{}ine the stage 1 EM corrections
\begin{equation}\hspace{2.5cm}
\Delta^1 a_{cc} = a_{cc}-\tilde{a}_{cc}\,\,\,\,,\,\,\,\, \Delta^1 a_{c0} = a_{c0}- \tilde{a}_{c0}\,\,.
\label{eq:aa}
\end{equation}
From Eqs.~(\ref{eq:m}), (\ref{eq:r}), (\ref{eq:z}) and (\ref{eq:aa}), we have
\begin{equation}\hspace{2.5cm}
\Delta^1 a_{cc} = +0.00078 \,\mathrm{fm}\,\,\,\,,\,\,\,\, \Delta^1 a_{c0} =+0.0030 \,\mathrm{fm}\,\,.
\label{eq:bb}
\end{equation}
These are the essential results from the present calculation.

It is instructive to look at the contributions of the three ef\mbox{}fects to the corrections $\Delta^1 a_{cc}$ and $\Delta^1 a_{c0} $. For 
$\Delta^1 a_{cc}$ we have
\begin{equation}\hspace{0cm}
\Delta a_{cc}(pc) = -0.00214 \,\mathrm{fm}\,,\Delta a_{cc}(md) =-0.00071 \,\mathrm{fm}\,,\Delta a_{cc}(\gamma n) =+0.00363\,\mathrm{fm} .
\label{eq:cc}
\end{equation}
The smallness of the total correction is thus seen to be the result of a near cancellation between the ef\mbox{}fect of the $\gamma n$ channel, which is 
large, and the combined ef\mbox{}fect of $V^{pc}$ and the mass dif\mbox{}ferences. One sees how important it is to estimate the ef\mbox{}fect of the 
$\gamma n$ channel as reliably as possible. For $\Delta^1 a_{c0} $ the situation is quite dif\mbox{}ferent. We now have
\begin{equation}\hspace{0cm}
\Delta a_{c0}(pc) = -0.0002 \,\mathrm{fm}\,,\Delta a_{c0}(md) =+0.0029 \,\mathrm{fm}\,,\Delta a_{c0}(\gamma n) =+0.0003\,\mathrm{fm} .
\label{eq:dd}
\end{equation}
The ef\mbox{}fect of the $\gamma n$ channel is now very small, and the correction is due almost entirely to the mass dif\mbox{}ferences.

Uncertainties in the corrections given in Eq.~(\ref{eq:bb}) arise from: (a) the shapes assumed for the components of $\widetilde{\mathbf{V}}_h$; (b) the 
errors of the f\mbox{}ive experimentally determined quantities which are f\mbox{}itted; (c) the approximation of using an isospin invariant form for 
$\widetilde{\mathbf{V}}_h$. The quantity $\Delta a_{cc}(pc)$ has a signif\mbox{}icant uncertainty due to lack of knowledge of the shapes of the components 
of $\widetilde{\mathbf{V}}_h$. This was estimated as in Refs.~\cite{E,P} by varying the range parameter between $0.8$ fm and $1.2$ fm and simultaneously 
adjusting the strengths of the potentials so that the f\mbox{}ive scattering parameters discussed earlier in this section remain constant. The quantity 
$\Delta a_{cc}(\gamma n)$ suf\mbox{}fers from the same uncertainty, and has in addition a signif\mbox{}icant error due to the uncertainty in 
$a_{c\gamma}$, 
$a_{0\gamma}$ and $a_{\gamma\gamma}$. The uncertainty in $\Delta a_{c0}(md)$, the dominant part of $\Delta^1 a_{c0}$, comes from uncertainties in 
$\widetilde{\mathbf{V}}_h$ (the shapes of $\widetilde{V}_{1/2}$ and $\widetilde{V}_{3/2}$ and the assumption of isospin invariance).

Unfortunately, in the situation under consideration one cannot achieve the precision associated with the errors quoted in the experimental results. The 
variations described in the previous paragraph can lead to rather rough estimates only. It is therefore wise to err on the side of caution and increase 
these estimates accordingly. We think that, on the basis of the estimates we have made, the errors on each of the components of the corrections could be 
as high as $20\%$. This leads to the following f\mbox{}inal values for the corrections
\begin{equation}\hspace{2cm}
\Delta^1 a_{cc} = +0.0008(8) \,\mathrm{fm}\,\,\,\,,\,\,\,\, \Delta^1 a_{c0} =+0.0030(6) \,\mathrm{fm}\,\,.
\label{eq:ee}
\end{equation}

Combining Eqs.~(\ref{eq:m}), (\ref{eq:r}) and (\ref{eq:ee}), we obtain the f\mbox{}inal results
\begin{equation}\hspace{2.5cm}
\tilde{a}_{cc}= 0.1214(8)\,\mathrm{fm}\,\,,\,\,\,\,  \tilde{a}_{c0}=-0.1757(21)\, \mathrm{fm}\,\,\,.
\label{eq:ff}
\end{equation}
(We have given results for the $s$-wave scattering parameters in fm because it is a precisely def\mbox{}ined unit. In units which are in common use, the 
f\mbox{}inal results in Eq.~(\ref{eq:ff}) are $\tilde{a}_{cc}=0.0859(6) \mu_c^{-1}$ and $\tilde{a}_{c0}=-0.1243(15) \mu_c^{-1}$.)

It is important to note that the error in $\tilde{a}_{cc}$ arises almost entirely from the uncertainty in $\Delta^1 a_{cc}$, so that an improvement in the 
accuracy of the measurement of $W(3p\rightarrow 1s)$ will not reduce the error in $\tilde{a}_{cc}$. On the other hand, almost all the error in 
$\tilde{a}_{c0}$ comes from the experimental uncertainty in $\Gamma$. If the error in the measurement of $\Gamma$ can be reduced to $1\%$, the uncertainty 
in $\tilde{a}_{c0}$ in Eq.~(\ref{eq:ff}) will be reduced from $(21)$ to $(11)$.

In the discussion following Eq.~(35) of Ref.~\cite{D}, we emphasised the very large dif\mbox{}ference between the values of $\tilde{a}_{cc}$ which come 
from pionic H and from our PSA of the combined $\pi^+p$ and $\pi^-p$ elastic scattering databases for $T_{\pi} \leq 100$ MeV. Converting from the unit 
$\mu_c^{-1}$ to fm, the latter gives the value $\tilde{a}_{cc}=0.1127(16)$ fm. The dif\mbox{}ference between this value and the value $0.1214(8)$ fm given 
in Eq.~(\ref{eq:ff}) is thus $0.0087(18)$ fm. This large dif\mbox{}ference is a subtle consequence of the violation of isospin invariance in the 
electromagnetically modif\mbox{}ied hadronic interaction, which is convincingly demonstrated in Section 7 of Ref.~\cite{D} and in Section 5 of this paper. 
The phase shift $\tilde{\delta}_{1/2}$ used in Ref.~\cite{D} to analyse $\pi^- p$ elastic scattering data is an artif\mbox{}icial construct, which enables 
the analysis to be done in a framework of formal isospin invariance. When isospin invariance is violated, $\tilde{\delta}_{1/2}$ goes to a nonzero value 
at threshold, so that the scattering length $\tilde{a}_{1/2}$ given in Ref.~\cite{D} is an artefact of the parameterisation of $\tilde{\delta}_{1/2}$, 
which forces it to behave like $q_c$ near threshold, and the extrapolation to threshold (using this artif\mbox{}icial parameterisation) of values of 
$\tilde{\delta}_{1/2}$ obtained from data at momenta $q_c \approx 80$ MeV/c upwards. Therefore the value of $\tilde{a}_{cc}$ extracted from pionic H 
cannot be compared directly with the value of $(2 \tilde{a}_{1/2} + \tilde{a}_+)/3$ obtained in the PSA of Ref.~\cite{D}.

\section{Comparison with other calculations of the electromagnetic corrections}

As mentioned already in the introduction, in our method of calculating the EM corrections to the pionic H data we use hadronic potentials which are of the 
same form as those of Refs.~\cite{E,P}. This implies that the EM corrections which were applied to the low energy $\pi^{\pm}p$ scattering data in our PSA 
of Ref.~\cite{D} are compatible with the ones obtained in the present paper for the pionic H case. For the PSA, it was suf\mbox{}f\mbox{}icient, in 
analysing data from $q_c \approx 80$ MeV/c upwards, to take the $\gamma n$ channel into account by using the perturbative result of 
Ref.~\cite{Nordita}. It has practically no ef\mbox{}fect on the results of the PSA. However, for our calculation at threshold, the ef\mbox{}fect of the 
$\gamma n$ channel turns out to be very signif\mbox{}icant in the case of the correction to $a_{cc}$, thus justifying the full three-channel calculation 
we have made. In these respects, our calculations of the EM corrections to the pionic H data dif\mbox{}fer considerably from the older work of 
Refs.~\cite{I,KG}, both of which use primitive forms of the hadronic potentials which reproduce only the scattering lengths, in contrast to our forms 
which also reproduce the phase shifts for $T_{\pi} \leq 100$ MeV. Furthermore, Ref.~\cite{KG} is a nonrelativistic calculation which uses much older 
values of the scattering lengths. In Ref.~\cite{I} the $\gamma n$ channel is taken into account in an unsatisfactory way, by using the Panofsky ratio to 
alter the input value of $a_{c0}$ in what remains a two-channel calculation, whereas we have included this channel in the most realistic way possible in a 
three-channel calculation. In addition, Refs.~\cite{I,KG} do not take account of the accurate relations given in Ref.~\cite{G}, between $\Delta W$ and 
$a_{cc}$ and between $\Gamma_0$ and $a_{c0}$. This means that our corrections and those of Ref.~\cite{I} dif\mbox{}fer by the factors in the parentheses 
of Eqs.~(\ref{eq:i},\ref{eq:p}). For these reasons, we consider that our calculations supersede the earlier potential model calculations in 
Refs.~\cite{I,KG}.

A quite dif\mbox{}ferent method of calculating the EM corrections is used in Ref.~\cite{ELW}. In a nonrelativistic approach using Coulomb wave functions, 
with a short-range hadronic interaction and extended charge distributions of $\pi^-$ and $p$, the authors identify four components of the corrections. Two 
are the contributions from the interference of $V^{vp}$ and $V^{ext}$ with the hadronic potential $\widetilde{V}$; the other two are a 'renormalisation' 
term and a 'gauge' term. The interference terms are obtained using an argument based on the change at the origin in the nonrelativistic $1s$ wave function 
arising from $V^{pc}$, when $V^{vp}$ and $V^{ext}$ are added to $V^{pc}$. This is a purely EM calculation, and the results are therefore claimed to be 
independent of $\widetilde{V}$. We can see no theoretical justif\mbox{}ication for this procedure of artif\mbox{}icially introducing the wave function at 
the origin as a replacement for the lowest Bohr energy in the DGBT formulae, and then calculating corrections via changes in this wave function. Moreover, 
while the shift in the $1s$ level position due to the interference of $V^{vp}$ with $\widetilde{V}$ happens to be correct ($-0.0034$ eV), that for 
$V^{ext}$ comes out to be $+0.0060$ eV, whereas the correct result is $+0.0034$ eV, which gives the cancellation reported in Section 2 and in 
Ref.~\cite{I}. The subthreshold calculation discussed in Section 2 shows that these contributions separately have a signif\mbox{}icant variation with the 
range of the hadronic potential, and we have given the error which results from this for the sum of the contributions. Ref.~\cite{ELW} also incorrectly 
says that the results given there for the interference contributions agree with those given in Ref.~\cite{I}.

While Ref.~\cite{ELW} has a discussion of the multichannel situation in Subsection 2.2, it does not explicitly give the terms which appear in parentheses 
in our Eqs.~(\ref{eq:i},\ref{eq:p}), nor are these (admittedly small) terms taken into account in their corrections. Further, we cannot see any 
corrections in Ref.~\cite{ELW} which correspond to the corrections due to external mass dif\mbox{}ferences appearing in Eqs.~(\ref{eq:cc},\ref{eq:dd}).

That leaves the renormalisation and gauge terms in Ref.~\cite{ELW}. Each of these terms involves an integral over the extended charge distributions, so it 
appears that some of the ef\mbox{}fect of extended charges is contained in these terms. In our approach, however, $V^{ext}$ is fully accounted for by its 
direct contribution to the position of the $1s$ level and the interference of $V^{ext}$ with $\widetilde{V}$. It also does not seem possible to see where 
the interference of $V^{pc}$ with $\widetilde{V}$, which gives the corrections designated (pc) in Eqs.~(\ref{eq:cc},\ref{eq:dd}), appears in the 
corrections of Ref.~\cite{ELW}. So here we deal with two very dif\mbox{}ferent approaches to the calculation of the EM corrections, with no discernible 
way of comparing them.

Further, the gauge terms in Ref.~\cite{ELW}, for the corrections to $a_{cc}$ and $a_{c0}$, are much more uncertain than is claimed there. They are 
proportional to the range parameters $b^h_{cc}$ and $b^h_{c0}$ respectively, which appear in the low energy expansions of $K_{cc}$ and $K_{c0}$. 
Ref.~\cite{ELW} uses for these quantities values given in Ref.~\cite{Hoehler}, taken from a PSA of $\pi N$ scattering data obtained before 1983. 
Practically none of these data were used in our recent PSA~\cite{D}. The numbers in Ref.~\cite{Hoehler} are no longer of any value. Moreover, as we 
explained at the end of Section 3, because of the violation of isospin invariance in the electromagnetically modif\mbox{}ied hadronic interaction, the PSA 
of Ref.~\cite{D} does not lead to a reliable estimate of $\tilde{a}_{cc}$. For $\tilde{b}_{cc}$ the situation is even more uncertain. The value obtained 
from the PSA cannot be accepted as a reliable determination; it is merely a rough guess based on phase shifts for $q_c \gtrsim 80$ MeV/c. The same holds 
for $\tilde{b}_{c0}$. It is not possible at present to obtain reliable values of the range parameters from a PSA. As a result, the values of the EM 
corrections proposed in Ref.~\cite{ELW} are quite uncertain.

In our calculations, the range parameters are replaced by potentials which, as explained in Refs.~\cite{E,P}, are constructed to reproduce the phase 
shifts for $q_c \gtrsim 80$ MeV/c. The lack of experimental data at very low energies increases the uncertainty in the potentials; it allows a wider 
interval of values of the gaussian range parameter $a$ which yield acceptable f\mbox{}its than would otherwise be the case. We have taken this uncertainty 
in $a$ into account in the estimates of the errors in our corrections given in Eq.~(\ref{eq:ee}).

In Ref.~\cite{EI} the dispersive EM correction to $a_{cc}$ which arises from the radiative processes $\pi^-p\rightarrow\gamma X$($X =n,\Delta$) is 
calculated in the heavy baryon limit. Table 1 of Ref.~\cite{EI} gives the correction $\Delta a_{cc} (\gamma n) = 0.00424(13)$, while our result 
(Eq.~(\ref{eq:cc})) is $0.00363(72)$ fm. These two results, which have been obtained in completely dif\mbox{}ferent ways, agree quite well. The case 
$X = \Delta$ is very dif\mbox{}ferent. In the potential model, the ef\mbox{}fect of a $\gamma \Delta$ intermediate state is a residual EM contribution to 
$\widetilde{V}$, and has to be calculated in another way. The result given in Table 1 of Ref.~\cite{EI} is $\Delta a_{cc} (\gamma \Delta) = + 0.00332$ fm. 
There is no other estimate of this correction. One should note that diagrams with a $\gamma \pi N$ intermediate state in principle appear in the 
calculation of Ref.~\cite{T}, which we shall discuss in a moment, but at a higher order in the perturbation expansion. An ef\mbox{}fect due to 
$\gamma \Delta$ as large as that given in Ref.~\cite{EI} would imply that higher orders of the expansion give contributions which are comparable in 
magnitude to the one-loop result given in Ref.~\cite{T}.

Gasser et al.~\cite{T} have made a calculation of the EM correction to $a_{cc}$ using baryon chiral perturbation theory (ChPT) in the infrared 
regularisation. The calculation was carried out up to diagrams containing one loop; it is a calculation of the complete EM correction to $a_{cc}$ to this 
order. An unambiguous splitting of this correction into stage 1 and stage 2 components is a very dif\mbox{}f\mbox{}icult task, and may not even be 
possible (for example, due to the ultraviolet divergences). Nevertheless we can say that some parts of the correction correspond to the ef\mbox{}fects 
calculated (to all orders) with our potential model. In addition, there are much more important parts of the correction of Ref.~\cite{T}, which are not 
taken into account in our calculation. They give a large negative correction which swamps the small correction $\Delta^1 a_{cc}$ in Eq.~(\ref{eq:ee}). 
Unfortunately, this correction is very uncertain, due to a lack of knowledge of the low energy constant $f_1$.

For the width, there is a calculation using baryon ChPT reported in Ref.~\cite{Zemp}, but only to lowest order (no loops). It contains contributions 
which are not accounted for in the potential model calculation; the total contribution is positive, but very small compared with the correction 
$\Delta^1 a_{c0}$ in Eq.~(\ref{eq:ee}).

\section{Violation of isospin invariance}

If the electromagnetically modif\mbox{}ied hadronic interaction were to satisfy isospin invariance, we would have the equality
\[
\hspace{3cm}
\tilde{a}_{cc}+\sqrt{2}\tilde{a}_{c0}=\tilde{a}_+\,\,\,,
\]
where $\tilde{a}_+$ is the $s$-wave scattering length for elastic $\pi^+ p$ scattering. The value of $\tilde{a}_+$ needs to be determined by means of a 
PSA of low energy $\pi^+ p$ scattering data, using EM corrections calculated in a manner consistent with the calculation of the corrections in 
Eq.~(\ref{eq:ee}). Our analysis \cite{D} of the $\pi^+ p$ scattering data for $T_{\pi} \leq 100$ MeV gives
\begin{equation}\hspace{4cm}
\tilde{a}_+= -0.1062(55)\,\mathrm{fm}\,\,.  
\label{eq:gg}
\end{equation}
When the analysis is made of a database which combines $\pi^+ p$ and $\pi^- p$ elastic scattering data up to $100$ MeV, using a more restrictive model for 
the f\mbox{}itting, the value of $\tilde{a}_+$ is practically unaltered, but the error decreases from (55) to (23). For the purpose of testing for isospin 
invariance, the larger uncertainty in Eq.~(\ref{eq:gg}) should be used.

From the results in Eq.~(\ref{eq:ff}) we have
\begin{equation}\hspace{2.5cm}
\tilde{a}_{cc}+\sqrt{2}\tilde{a}_{c0}= -0.1271(31)\,\mathrm{fm}\,\,.
\label{eq:hh}
\end{equation}
Almost all the error in Eq.~(\ref{eq:hh}) arises from the uncertainty in $\tilde{a}_{c0}$. Reducing the error in $\Gamma$ to $1\%$ would reduce the 
uncertainty in Eq.~(\ref{eq:hh}) to (17). The dif\mbox{}ference between the numbers in Eqs.~(\ref{eq:gg},\ref{eq:hh}) is $0.0209(63)$ fm, which amounts 
to $3.3$ standard deviations. This is evidence for a signif\mbox{}icant violation of isospin invarance at the electromagnetically modif\mbox{}ied hadronic 
level. If the experiment \cite{B} achieves its hoped for accuracy for the measurement of $\Gamma$, the error in the dif\mbox{}ference will change from 
(63) to (58). If the result for $\Gamma$ remains approximately the same, the dif\mbox{}ference will become even more signif\mbox{}icant.

This result strengthens the evidence for the violation of isospin invariance at the electromagnetically modif\mbox{}ied hadronic level given in 
Ref.~\cite{D}, which comes from the inability of the phase shifts obtained from the analysis of the combined $\pi^{\pm} p$ elastic scattering data to 
reproduce the data on the charge exchange reaction $\pi^- p \rightarrow \pi^0 n$.

\begin{ack}
We are grateful to Detlev Gotta for communicating to us the current results of the PSI pionic hydrogen experiment, and to J{\"u}rg Gasser and Akaki 
Rusetsky for their helpful answers to our questions concerning the work of the Bern group. We would also like to thank the two reviewers of our paper for 
their useful recommendations and constructive criticism.
\end{ack}

\end{document}